\documentclass{pasj00}
\draft

\begin{document}
\SetRunningHead{M. Omiya et al.}{A Massive Substellar Companion to HD 119445}
\Received{0000/00/00}
\Accepted{0000/00/00}

\title{A Massive Substellar Companion to the Massive Giant HD 119445}

\author{
Masashi \textsc{Omiya},\altaffilmark{1}
Hideyuki \textsc{Izumiura},\altaffilmark{2,3}
Inwoo \textsc{Han},\altaffilmark{4}
Byeong-Cheol \textsc{Lee},\altaffilmark{4,5}

Bun'ei \textsc{Sato},\altaffilmark{6}
Eiji \textsc{Kambe},\altaffilmark{2}
Kang-Min \textsc{Kim},\altaffilmark{4}
Tae Seog \textsc{Yoon},\altaffilmark{5}
Michitoshi \textsc{Yoshida},\altaffilmark{2}

Seiji \textsc{Masuda},\altaffilmark{7}
Eri \textsc{Toyota},\altaffilmark{8}
Seitaro \textsc{Urakawa},\altaffilmark{9} and
Masahide \textsc{Takada-Hidai},\altaffilmark{10}
}

\altaffiltext{1}{Department of Physics, Tokai University, 1117 Kitakaname, Hiratsuka, Kanagawa 259-1292, Japan\\ohmiya@peacock.rh.u-tokai.ac.jp}
\altaffiltext{2}{Okayama Astrophysical Observatory, National Astronomical Observatory of Japan, Asakuchi, Okayama 719-0232, Japan}
\altaffiltext{3}{Department of Astronomical Science, The Graduate University for Advanced Studies, Shonan Village, Hayama, Kanagawa 240-0193, Japan}

\altaffiltext{4}{Korea Astronomy and Space Science Institute, 61-1 Whaam-dong, Youseong-gu, Taejeon 305-348, South Korea}
\altaffiltext{5}{Department of Astronomy and Atmospheric Sciences, Kyungpook National University, Daegu 702-701, South Korea}

\altaffiltext{6}{Global Edge Institute, Tokyo Institute of Technology, 2-12-1-S6-6 Ookayama, Meguro-ku, Tokyo 152-8550, Japan}

\altaffiltext{7}{Tokushima Science Museum, Asutamu Land Tokushima, Itano-gun, Tokushima 779-0111, Japan}
\altaffiltext{8}{Kobe Science Museum, 7-7-6 Minatojimanakamachi, Chuo-ku, Kobe, Hyogo 650-0046, Japan}
\altaffiltext{9}{Bisei Spaceguard Center, Japan Spaceguard Association, 1716-3 Okura, Bisei-cho, Ibara, Okayama 714-1411, Japan}
\altaffiltext{10}{Liberal Arts Education Center, Tokai University, 1117 Kitakaname, Hiratsuka, Kanagawa 259-1292, Japan}




\maketitle

\begin{abstract}
We detected a brown dwarf-mass companion around the intermediate-mass giant star HD 119445 (G6III) using the Doppler technique. This discovery is the first result from a Korean$-$Japanese planet search program based on precise radial velocity measurements. The radial velocity of this star exhibits a periodic Keplerian variation with a period, semi-amplitude and eccentricity of 410.2 days, 413.5 m s$^{-1}$ and 0.082, respectively. Adopting a stellar mass of 3.9 $M_{\odot}$, we were able to confirm the presence of a massive substellar companion with a semimajor axis of 1.71 AU and a minimum mass of 37.6 $M_{\mathrm{J}}$, which falls in the middle of the brown dwarf-mass region. This substellar companion is the most massive ever discovered within 3 AU of a central intermediate-mass star. The host star also ranks among the most massive stars with substellar companions ever detected by the Doppler technique. This result supports the current view of substellar systems that more massive substellar companions tend to exist around more massive stars, and may further constrain substellar system formation mechanisms.
\end{abstract}

\section{Introduction}
A brown dwarf is defined as an object that has a mass between the deuterium-burning limit of $\sim$ 13 $M_{\mathrm{J}}$ and the hydrogen-burning limit of $\sim$ 80 $M_{\mathrm{J}}$ (e.g., \cite{Burrows1997}). So far, brown dwarf-mass companions to normal stars have been searched for by various techniques, such as precise Doppler measurements (e.g., \cite{Nidever2002}; \cite{Patel2007}), direct imaging (e.g., \cite{McCarthy2004}; \cite{Lafreniere2007}), spectroscopic (e.g., \cite{Neuhauser2004}) and astrometric observations (e.g., \cite{Halbwachs2000}; \cite{Zucker2001}). Although some brown dwarf-mass companions have been detected by these methods, one intriguing result from these observations is that brown dwarf-mass companions close ($\lesssim$ 1000 AU) to solar-type stars are conspicuously scarce compared to planetary and stellar companions (e.g., \cite{Gizis2001}; \cite{McCarthy2004}). From the combined results of various precise Doppler surveys of nearby FGK-type stars, \citet{Grether2006} reported that less than 1 \% of solar-type stars harbor brown dwarf-mass companions with orbital periods less than 5 years, while 11 $\pm$ 3 \% and 5 $\pm$ 2 \% have stellar companions and giant planets, respectively. Moreover, according to a near-infrared imaging survey of the Sco OB2 association by \citet{Kouwenhoven2007}, the frequency of brown dwarf-mass companions ($\gtrsim$ 30 $M_{\mathrm{J}}$) with orbital separations of 130$-$520 AU around intermediate-mass A or late-B type stars is 0.5 $\pm$ 0.5 \%. This rate is lower than that of stellar companions by one order of magnitude. The paucity of brown dwarf-mass companions relative to both planetary companions and stellar companions is known as a ``brown dwarf desert'' (e.g., \cite{Marcy2000}; \cite{Grether2006}). The existence of the desert may be explained by orbital migration \citep{Armitage2002} or ejection \citep{Reipurth2001} of brown dwarf-mass companions, or may suggest a bimodal mass function of substellar companions that could be produced by two distinct formation mechanisms: core-accretion in protoplanetary disks (e.g., \cite{Ida2004}; \cite{Alibert2005}), forming mostly planetary companions, and gravitational fragmentation in molecular clouds or protoplanetary disks (e.g., \cite{Boss1997}; \cite{Bate2000}; \cite{Rice2003}; \cite{Stamatellos2009}), forming mostly stellar or brown dwarf-mass companions.

To date, several precise Doppler surveys of subgiants, giants and early-type dwarfs have uncovered four brown dwarf-mass companions (13 $M_{\mathrm{J}}$ $<$ $M_{\mathrm{2}} \mathrm{sin} i_{p}$ $\leq$ 25 $M_{\mathrm{J}}$) and 21 planetary companions (0.6 $M_{\mathrm{J}}$ $\leq$ $M_{\mathrm{2}} \mathrm{sin} i_{p}$ $\leq$ 13 $M_{\mathrm{J}}$) around intermediate-mass stars (1.5 $M_{\mathrm{\odot}}$ $\leq$ $M$ $\leq$ 5 $M_{\odot}$) (\cite{Sato2003}; \cite{Hatzes2005}; \cite{Setiawan2005}; \cite{Galland2006}; \cite{Hatzes2006}; \cite{Johnson2007}; \cite{Lovis2007}; \cite{Niedzielski2007}; \cite{Robinson2007}; \cite{Sato2007}; \cite{Johnson2008}; \cite{Liu2008}, \cite{Sato2008a}, \yearcite{Sato2008b}; \cite{Liu2009}; \cite{Niedzielski2009}). Preliminary statistics suggest that some properties of substellar systems orbiting intermediate-mass stars are not necessarily similar to those of solar-type stars (e.g., \cite{Butler2006}). The masses of substellar companions orbiting intermediate-mass stars tend to be larger than those around solar-type stars (\cite{Lovis2007}; \cite{Hekker2008}). The orbital semimajor axes of all substellar companions detected around intermediate-mass giants are larger than about 0.6 AU (\cite{Johnson2007}; \cite{Sato2008a}; \cite{Niedzielski2009}). In comparison, substellar companions of solar-type stars can also be found in much closer orbits ($\geq$ 0.017 AU). The metallicity [Fe/H] of many planet-harboring intermediate-mass stars is lower than that typically observed for solar-type stars with planetary companions (\cite{Fischer2005}; \cite{Pasquini2007}; \cite{Takeda2008}). Additional comparisons between planetary systems orbiting intermediate-mass stars and other types of stars would be of great interest for better understanding the dependence of substellar system formation on the central star mass.

We report the discovery of a brown dwarf-mass companion orbiting the intermediate-mass giant HD 119445. This is the first result of an ongoing Korean$-$Japanese planet search program carried out at Bohyunsan Optical Astronomy Observatory (BOAO, Korea) and Okayama Astrophysical Observatory (OAO, Japan). The planet search program is introduced in section 2. We describe the properties of the host star and the orbital motion in sections 3 and 4. The cause of the radial velocity variation and an upper limit on the companion mass are discussed in section 5. In section 6, we consider the implications of this discovery for the current picture of substellar companions.

\section{Korean$-$Japanese Planet Search Program}
In 2005, we started a joint planet search program between Korean and Japanese researchers to search for planets around GK-type giant stars using a precise Doppler technique with using the 1.8-m telescope at BOAO and the 1.88-m telescope at OAO. This survey program is an extended version of the ongoing OAO planet search program \citep{Sato2005} and part of an international collaboration among researchers from Korea, China and Japan (an East-Asian Planet Search Network, EAPS-Net; \cite{Izumiura2005}). The collaboration aims at clarifying the properties of planetary systems around intermediate-mass stars by surveying more than 800 GK giants for planets at OAO, BOAO, the Xinglong station (China), and the Subaru Telescope.

For the Korean$-$Japanese planet search program, we selected about 190 target stars from the $Hipparcos$ catalog based on the following criteria: color-index 0.6 $<$ $B-V$ $<$ 1.0, absolute magnitude $-$3 $<$ $M_{v}$ $<$ 2, declination $\delta$ $>$ $-$25 $^{\circ}$, and visual magnitude 6.2 $<$ $V$ $<$ 6.5. These targets are fainter than those of the OAO and Xinglong program (\cite{Sato2005}; \cite{Liu2008}). We divided the targets into two parts: one for BOAO and the other for OAO. Each is observed independently at the assigned observatory, although a star that exhibits a large radial velocity variation is observed intensively at both observatories.

We will also carry out abundance analyses for all the target stars to derive fundamental stellar parameters and chemical compositions, and to investigate correlations between these stellar characteristics and the existence of orbiting planets.

\subsection{BOES Observations}
Radial velocity observations at BOAO were carried out with the 1.8-m telescope and BOAO Echelle Spectrograph (BOES; \cite{Kim2007}), a fiber-fed high resolution echelle spectrograph. For precise radial velocity measurements, we placed an iodine absorption cell (I$_{2}$ cell) in the optical path in front of the fiber entrance of the spectrograph \citep{Kim2002} and used a 200-$\mu$m fiber, obtaining a wavelength resolution $R$ = $\lambda$/$\Delta{\lambda}$ $\sim$ 51000. The spectra covered a wavelength range of 3500$-$10500 $\mathrm{\AA}$. We used the range between 5000 $\mathrm{\AA}$ and 5900 $\mathrm{\AA}$, a region covered by many I$_{2}$ absorption lines, for precise radial velocity measurements. We also made use of Ca II H lines at around 3970 $\mathrm{\AA}$ as chromospheric activity diagnostics. Echelle data reduction was performed using the IRAF\footnote[1]{IRAF is distributed by the National Optical Astronomy Observatories, which is operated by the Association of Universities for Research in Astronomy, Inc. under cooperative agreement with the National Science Foundation, USA.} software package in the standard manner.

\begin{figure}
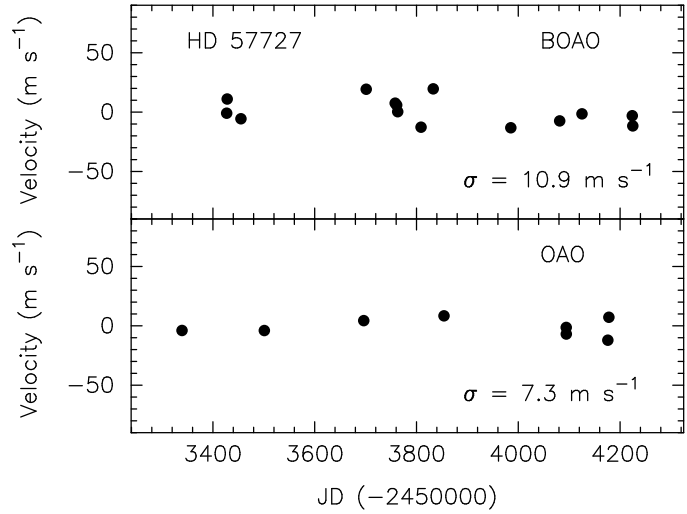

  \begin{center}
    \FigureFile(90mm,90mm){figure1.eps}
  \end{center}
  \caption{Radial velocity variations of radial velocity standard star HD 57727 observed at BOAO (upper) and OAO (lower).}
\label{fig:RVs}
\end{figure}

Precise radial velocities for the BOES data were derived using a modeling method detailed in \citet{Sato2002}, based on the method of \citet{Butler1996} and improved and optimized for BOES data analysis. We used seven Gaussian profiles with a common full width half maximum of 1.3 pixels, placed at 1 pixel intervals to reconstruct the BOES instrumental profile, and a fourth-order Legendre polynomial to describe the wavelength scale. The number of Gaussians, the width and interval of the Gaussian, and the order of the Legendre polynomial were chosen to minimize the long-term radial velocity dispersion of standard star HD 57727, which is known to have a small radial velocity scatter of $\lesssim$ 7 m s$^{-1}$ based on OAO observations over 7 years. We employed the extraction method described in \citet{Sato2002} to prepare a stellar template spectrum from some stellar spectra taken through the I$_{2}$ cell (I$_{2}$+stellar spectra). Stellar radial velocity was determined by averaging radial velocities from each of $\sim$ 200 spectral segments (each $\sim$ 5 $\mathrm{\AA}$ long) except those with the worst least-squares fit. This technique allowed us to achieve a Doppler precision of $\sim$ 11 m s$^{-1}$ over 2.3 years during this project (see figure \ref{fig:RVs}).

\subsection{HIDES Observations}
Radial velocity observations at OAO were carried out with the 1.88-m telescope and HIgh Dispersion Echelle Spectrograph (HIDES; \cite{Izumiura1999}) attached to the coud\'e focus of the telescope. For radial velocity measurements, we observed over the 5000 to 6200 $\mathrm{\AA}$ wavelength range with a slit width of 200 $\mu$m (0.76") giving a spectral resolution of 63000. An I$_{2}$ cell \citep{Kambe2002} was used for precise wavelength calibration. Echelle data reduction was performed using the IRAF software package in the standard manner. Stellar radial velocities were derived from the I$_{2}$-superposed stellar spectrum modeling technique detailed in \citet{Sato2002}, and gave a Doppler precision of $\sim$ 7 m s$^{-1}$ over 2.3 years during our program (see figure \ref{fig:RVs}). We also obtained stellar spectra without the I$_{2}$ cell using the same spectrograph setting for abundance analysis.

\section{ HD 119445 Stellar Parameters}
HD 119445 (HR 5160, HIP 66892) is located at 289 pc from the Sun according to the $Hipparcos$ parallax of $\pi$ = 3.46 $\pm$ 0.71 mas. The star is classified as a G6III giant star with $V$ = 6.30 and $B-V$ = 0.879 $\pm$ 0.004 \citep{ESA1997}. We derived an effective temperature $T_\mathrm{eff}$ of the star as $T_\mathrm{eff}$ = 5083 $\pm$ 103 K using a $(B-V)-T_{\mathrm{eff}}$ calibration of Alonso et al. (\yearcite{Alonso1999}, \yearcite{Alonso2001}). A luminosity of $L$ = 251 $\pm$ 95 $L_{\odot}$ was obtained from the absolute magnitude $M_{v}$ = $-$1.03 and the bolometric correction $B.C.$ = $-$0.23 based on the calibration of \citet{Alonso1999}. A stellar mass of $M$ = 3.9 $\pm$ 0.4 $M_{\odot}$ was estimated by interpolating the evolutionary tracks of \citet{Girardi2000} with the estimated $T_{\mathrm{eff}}$ and $L$. We determined the surface gravity to be log $g$ = 2.40 $\pm$ 0.17 and the stellar radius $R$ = 20.5 $\pm$ 9.3 $R_{\odot}$ from $M$, $L$ and $T_{\mathrm{eff}}$. A microturbulent velocity $V_{t}$ = 1.49 $\pm$ 0.20 km s$^{-1}$ and [Fe/H] = 0.04 $\pm$ 0.18 were derived from abundance analysis with a model atmosphere \citep{Kurucz1993} using equivalent widths of Fe I and Fe II lines measured from an I$_{2}$-free spectrum of HD 119445. We adopted gf-values of Fe I and Fe II lines from \citet{Takeda2005}. The stellar rotational velocity $v \mathrm{sin} i_{s}$ was found by \citet{Gray1989} and \citet{deMedeiros1999} to be 6.0 $\pm$ 0.6 km s$^{-1}$ and 6.9 $\pm$ 1.0 km s$^{-1}$, respectively. These values are larger than rotational velocities of most of late G-type giants. The stellar parameters are summarized in table \ref{Tab:SP}.

\begin{figure}
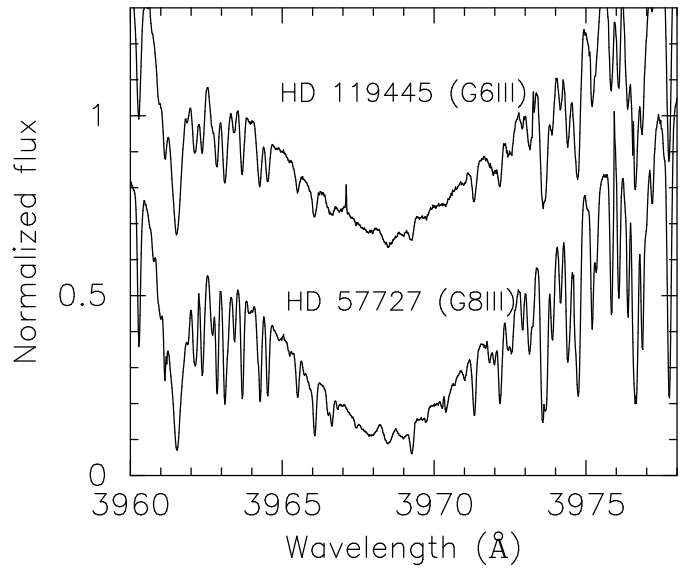

  \begin{center}
    \FigureFile(90mm,99mm){figure2.eps}
  \end{center}
  \caption{The spectral region around Ca II H lines of HD 119445 and HD 57727. The Ca II H line of HD 119445 does not seem to exhibit high activity, similar to the line of HD 57727.}
\label{fig:cah}
\end{figure}

Figure \ref{fig:cah} shows Ca II H lines of HD 119445 and the radial velocity standard star HD 57727. There is a lack of significant emission in the Ca II H line core of HD 119445 and HD 57727, which suggests chromospheric inactivity, although the correlation between chromospheric activity and intrinsic radial velocity jitter for giant stars is not yet well established. Moreover, $Hipparcos$ photometry collected from 187 observations of the star demonstrates the photometric stability of HD 119445 down to $\sigma$ $\sim$ 0.008 mag, which also suggests chromospheric inactivity for the star.

\section{Orbital Solution}
\begin{figure}
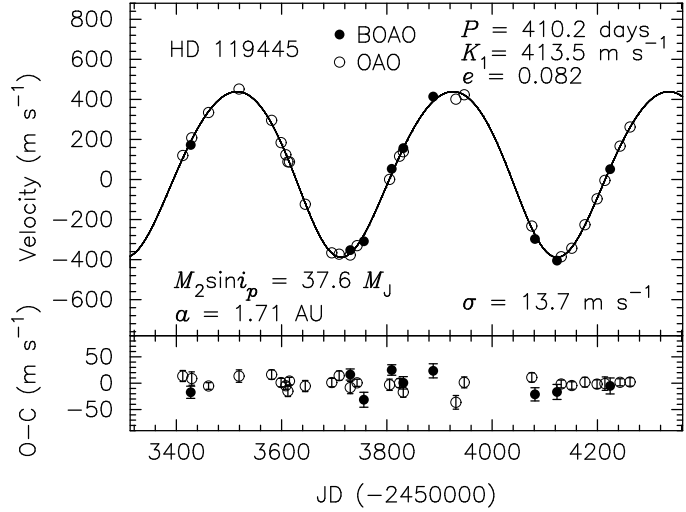

  \begin{center}
    \FigureFile(90mm,90mm){figure3.eps}
  \end{center}
  \caption{Upper panel: radial velocities of HD 119445 observed at BOAO ($filled$ $circles$) and OAO ($open$ $circles$). The Keplerian orbital curve we determined is shown by the solid line. Lower panel: Residuals to the best Keplerian fit.}
  \label{fig:RVofSS}
\end{figure}

We monitored the radial velocity of HD 119445 for 2.3 years from the beginning of the survey at both observatories. We accumulated 9 BOAO data points with a typical signal-to-noise ratio (S/N) of 180 pixel$^{-1}$, given an exposure time of 900$-$1200 s, and 27 OAO data points with a typical S/N of 140 pixel$^{-1}$, given an exposure time of 1200$-$1800 s. The observed radial velocities of HD 119445 are shown in figure \ref{fig:RVofSS} and listed in tables \ref{tab:RVvBOAO} (BOAO) and \ref{tab:RVvOAO} (OAO), together with estimated uncertainties. The best-fit Keplerian orbit derived from both the BOAO and OAO data has a period $P$ $=$ 410.2 $\pm$ 0.6 days, a velocity semi-amplitude $K_{1}$ $=$ 413.5 $\pm$ 2.6 m s$^{-1}$, and an eccentricity $e$ $=$ 0.082 $\pm$ 0.007. The best-fit curve is shown in figure \ref{fig:RVofSS} as a solid line overlaid on the observed velocities. We applied an offset of $\Delta$RV, $-$43.0 $\pm$ 5.1 m s$^{-1}$,  estimated concurrently with the orbital fit, to the BOAO data points by minimizing $\chi^{2}$ in fitting a Keplerian model to the combined BOAO and OAO velocity data. The difference in the velocity scales derived from the BOES and HIDES stellar templates yielded the offset. The best-fit parameters and uncertainties are listed in table \ref{tab:OPstar}. The uncertainties were estimated by a Monte Carlo approach.

The rms scatter of the residuals to the best-fit is 13.7 m s$^{-1}$. This value is comparable to the typical radial velocity scatter (10$-$20 m s$^{-1}$) of the G-type giants \citep{Sato2005} within the typical measurement errors of $\sim$ 13 m s$^{-1}$ (BOAO) and $\sim$ 9 m s$^{-1}$ (OAO). Since we did not find any significant periodic variation due to a second companion in the residuals, we conclude that HD 119445 would have no additional substellar companion with a period less than 800 days. Adopting a stellar mass $M$ = 3.9 $\pm$ 0.4 $M_{\odot}$ for HD 119445, we obtained a semimajor axis $a$ = 1.71 $\pm$ 0.06 AU and a minimum mass $M_{\mathrm{2}} \mathrm{sin} i_{p}$ = 37.6 $\pm$ 2.6 $M_{\mathrm{J}}$ for a companion. The error of the mass arose mostly from the stellar mass error of the host star.

\section{Line Shape Analyses and an Upper Limit of a Mass}
\begin{figure}
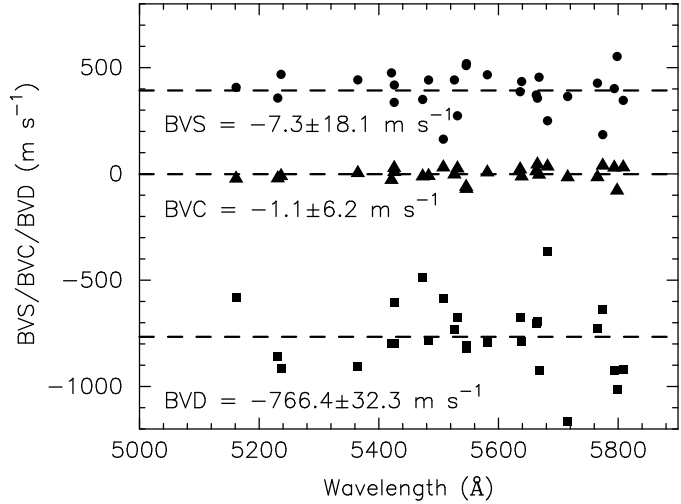

  \begin{center}
    \FigureFile(90mm,90mm){figure4.eps}
  \end{center}
  \caption{Bisector quantities calculated by cross-correlation profiles between the stellar templates of HD 119445 at peak (300$-$400 m s$^{-1}$) and valley ($-$400m s$^{-1}$) phases of the radial velocities. Shown are the bisector velocity span with an offset of 400 m s$^{-1}$ (BVS, $circles$), bisector velocity curvature (BVC, $triangles$) and bisector velocity displacement (BVD, $squares$). Average values and standard errors are shown in the figure. Dashed lines represent mean values.}
  \label{fig:bvs}
\end{figure}

The spectral-line shape analyses were performed using techniques described in \citet{Sato2007} to investigate other causes of the apparent radial velocity variation, such as rotational modulation and pulsation. For the analyses, we used two high-resolution stellar templates extracted from I$_{2}$+stellar spectra obtained at OAO. One template was constructed from four spectra with observed radial velocities ranging from 330 to 430 m s$^{-1}$ (peak), and the other from four spectra of around $-$370 m s$^{-1}$ (valley). Cross-correlation profiles of the templates were calculated for 27 spectral segments (4- to 5-\AA \ width each) that did not include severely blended lines or broad lines. Three bisector quantities were calculated for the cross-correlation profile of each segment: the velocity span (BVS), which is the velocity difference between two flux levels of the bisector; the velocity curvature (BVC), which is the difference of the velocity span of the upper half and lower half of the bisector; and the velocity displacement (BVD), which is the average of the bisector at three different flux levels. We used flux levels of 25\%, 50\%, and 75\% of the cross-correlation profile to calculate the above quantities. These bisector quantities for HD 119445 are shown in figure \ref{fig:bvs}. As expected under the planetary hypothesis, both the BVS and BVC (each average to $-$7.3 $\pm$ 18.1 m s$^{-1}$ and $-$1.1 $\pm$ 6.2 m s$^{-1}$, respectively) are essentially identical to zero, meaning that the cross-correlation profiles are symmetric. The dispersions of the BVS and BVD are relatively large. This is probably due to the broad absorption lines in the stellar spectra caused by the large rotational velocity. However, the average value of the BVD ($-$766.4 $\pm$ 32.3 m s$^{-1}$) is consistent with the velocity difference between the two templates. The value is more than 20 times larger than the dispersions; thus, the velocity difference between the templates is considered to be due to a parallel shift of spectral lines caused by orbital motion, not to variations in spectral line shapes. Hence, the observed radial velocity variation of HD 119445 is best explained by orbital motion of a companion  not by intrinsic activity such as rotational modulation or pulsation.

If we assume the orbit is randomly oriented, a 12 \% chance exists that the true mass exceeds the brown dwarf-mass limit of 80 $M_{\mathrm{J}}$ ($i_{p}$ $\lesssim$ 28$^{\circ}$). Assuming that the orbit of the companion is coplanar with the stellar equatorial plane, a small orbital inclination of less than 28$^{\circ}$ implies a stellar rotational velocity larger than 12 km s$^{-1}$. Single G-type giants with such a high rotational velocity are rare and would be identified as X-ray sources in $ROSAT$ observations. However, no X-ray emissions from HD 119445 have been detected \citep{Hunsch1998}, which suggests that the rotational velocity of HD 119445 is not fast and that the orbital inclination $i_{p}$ is not so small. Thus, the true mass of the companion may be smaller than 80 $M_{\mathrm{J}}$, which is the upper-limit mass for a brown dwarf.

\section{Discussion}
We detected a brown dwarf-mass companion orbiting the intermediate-mass giant star HD 119445. This discovery is the first result from our Korean$-$Japanese planet search program. The host star HD 119445 has a mass of 3.9 $M_{\odot}$. It is one of the most massive stars hosting substellar companions. HD 119445 b is the fifth brown dwarf-mass companion with a semimajor axis of less than 3 AU and the most massive brown dwarf-mass companion among those discovered around intermediate-mass stars. Now we found two brown dwarf-mass companions and ten planetary companions from the surveys at OAO, Xinglong and BOAO from a total of about 500 targets of the surveys (\cite{Sato2003}, \yearcite{Sato2007}, \yearcite{Sato2008a}; \yearcite{Sato2008b}; \cite{Liu2008}, \yearcite{Liu2009}; this work). The ratio between the numbers of brown dwarf-mass companions and planetary companions detected from the surveys of our GK-type giants seems comparable to that ($\leq$ 1 \% to $\sim$ 5 \%; \cite{Grether2006}) for solar-type stars, although the surveys are not yet complete and detection limits differ between solar-type and giant stars. This result may support the existence of a brown dwarf desert, a deficit of brown dwarf-mass companions relative to planetary companions, around intermediate-mass stars inside $a$ $\sim$ 3 AU orbital separation.

\begin{figure}
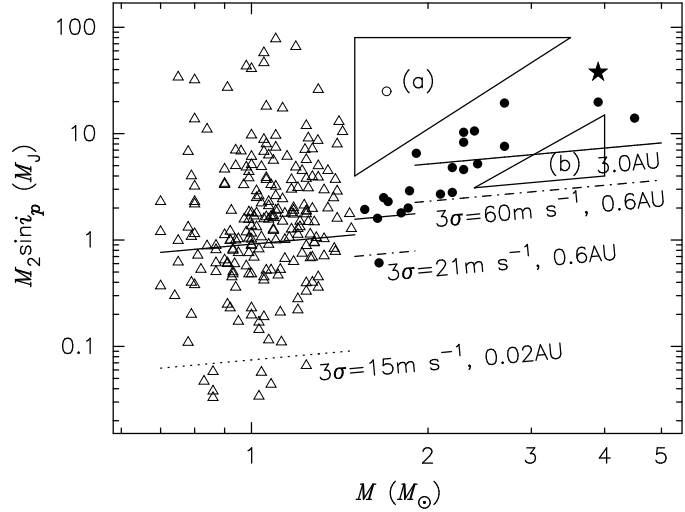

  \begin{center}
    \FigureFile(90mm,90mm){figure5.eps}
  \end{center}
  \caption{Primary star masses versus masses of substellar companions orbiting within 3 AU. $Open$ $triangles$, $filled$ $circles$ and an $open$ $circle$ represent companions orbiting solar-mass stars, intermediate-mass evolved stars (subgiants and giants) and an intermediate-mass dwarf (A-type star; HD 180777), respectively. A $star$ represents the HD 119445 system. Solid lines indicate the detection limits for the mass of companions orbiting at 3 AU, corresponding to three times of typical radial velocity jitters $\sigma$ of 5 m s$^{-1}$ for solar-mass stars (0.7 $M_{\odot}$ $\leq$ $M$ $<$ 1.5 $M_{\odot}$), 7 m s$^{-1}$ for  intermediate-mass subgiants (1.5 $M_{\odot}$ $\leq$ $M$ $\leq$ 1.9 $M_{\odot}$) and 20 m s$^{-1}$ for  intermediate-mass giants (1.9 $M_{\odot}$ $<$ $M$ $\leq$ 5 $M_{\odot}$). Dotted and dot-dashed lines indicate the detection limits for companions at 0.02 AU and 0.6 AU in solar mass stars and intermediate-mass evolved stars, respectively. Two regions devoid of substellar companions are denoted by (a) and (b).}
\label{fig:dmass}
\end{figure}

However, the existence of the desert may depend on host star's mass. In figure \ref{fig:dmass}, we plot masses of the companions detected within semi-major axis of 3 AU by precise Doppler surveys against their host star's masses; solar-mass stars (0.7 $M_{\odot}$ $\leq$ $M$ $<$ 1.5 $M_{\odot}$, $open$ $triangles$), intermediate-mass subgiants and giants (1.5 $M_{\odot}$ $\leq$ $M$ $\leq$ 5 $M_{\odot}$, $filled$ $circles$), an intermediate-mass dwarf (A-type star HD 180777, $open$ $circle$), and HD 119445 ($star$) ($The$  $Extrasolar$ $Planets$  $Encyclopadia$\footnote[2]{http://exoplanet.eu/, Version of 24/April/2008}; \cite{Halbwachs2000}; \cite{Tinney2001}; \cite{Nidever2002};  \cite{Vogt2002}; \cite{Endl2004}; \cite{Galland2006}; \cite{Liu2008}; \cite{Sato2008a}, \yearcite{Sato2008b}; this work). The detectable companion mass for a given host star mass depends on the orbital separation of its companion and the radial velocity jitter of the host star. Assuming that typical radial velocity jitters $\sigma$ for solar-mass stars, intermediate-mass subgiants (1.5$-$1.9 $M_{\odot}$) and giants (1.9$-$5 $M_{\odot}$) are $\sim$ 5 m s$^{-1}$, $\sim$ 7 m s$^{-1}$ and $\sim$ 20 m s$^{-1}$, respectively, we estimated the lower limits of companion masses detectable by precise Doppler surveys around a solar-mass star and intermediate-mass subgiant and giant at 3 AU (sold lines in figure \ref{fig:dmass}), corresponding to companion masses that provide the amplitude of three times of typical radial velocity jitters. We also indicate detectable masses for these stars at 0.02 AU (dotted lines) and 0.6 AU (dot-dashed lines), corresponding to the semimajor axes of the known innermost planets orbiting solar-type and intermediate-mass evolved stars.

Two unpopulated regions of substellar companions orbiting intermediate-mass subgiants and giants seem to exist in region (a) and (b)\footnote[3]{We exclude a brown dwarf-mass companion orbiting a possible high mass giant HD 13189 ($M$ $=$ 4.5 $\pm$ 2.5 $M_{\odot}$; \cite{Hatzes2005})  from the following discussion because of the large uncertainty in its host star's mass.}. A possible host star-companion mass correlation considered from unpopulated region (a) and (b) supports the current view that more massive substellar companions tend to exist around more massive stars, that are derived from the results of planet searches around various mass stars (\cite{Lovis2007}; \cite{Hekker2008}). 

All of the brown dwarf-mass companions to intermediate-mass evolved stars were found around those with $\geq$ 2.7 $M_{\odot}$ and there seems to be a paucity of such companions around those with 1.5$-$2.7 $M_{\odot}$ (region (a) in figure \ref{fig:dmass}). A brown dwarf-mass companion orbiting the A-type dwarf HD 180777 \citep{Galland2006} is regarded as a unique and interesting one since it populates the region (a). Such early-type dwarfs have larger typical radial velocity jitter $\sigma$ ($\geq$ 66 m s$^{-1}$ for late A-type dwarfs; \cite{Lagrange2009}) than that of evolved stars due to more rapid rotation and pulsation, and thus the planet mass distribution in early-type dwarfs is less well defined than that of intermediate-mass evolved stars. Considering the smaller number of survey targets of $\geq$ 2.7 $M_{\odot}$ (e.g., 35 \% of the 300 OAO targets; \cite{Takeda2008}) compared with that of 1.5$-$2.7 $M_{\odot}$, frequency of brown dwarf companions may become higher as stellar mass increases. This might favor gravitational instability in protostellar disks \citep{Rice2003} rather than fragmentation of proto-stellar clouds \citep{Bate2000} as the formation mechanism of brown dwarf-mass companions because stellar systems with larger difference in mass between primary and secondary stars are more difficult to form by the latter mechanism \citep{Bate2000}.

Also, there seems to be a possible paucity of lower-mass companions around 2.4$-$4 $M_{\odot}$ stars (region (b) in figure \ref{fig:dmass}). Although it is basically difficult to detect planets around such "noisy" stars with large intrinsic radial velocity variability ($\sigma$ $\sim$ 20 m s$^{-1}$), planets with mass $\geq$ 2.6$-$3.3 $M_{\mathrm{J}}$ ($\geq$ 5.7$-$7.4 $M_{\mathrm{J}}$) and $a$ $=$ 0.6 AU ($a$ $=$ 3.0 AU) should be above the current detection limit (3 $\sigma$ $\sim$ 60 m s$^{-1}$). Recently, \citet{Kennedy2008} predicted that the frequency of giant planets has a peak near 3 $M_{\odot}$ stars based on a core accretion scenario taking account of the movement of snow line along the evolution of accretion and the central stars. Moreover, if a formation mechanism works that invokes capturing of solid bodies migrating inward at the inner edge of the inactive magnetorotational instability-dead zone inside of the protoplanetary disk, gas giant planets could be formed efficiently at around 1 AU around intermediate-mass stars before the planetary disks deplete \citep{Kretke2009}. Increasing the number of known massive planetary companions around massive intermediate-mass stars by further radial velocity surveys would be of great interest to understand the formation mechanisms of giant planets around intermediate-mass stars.

\bigskip
This research was supported as a Korea-Japan Joint Research Project under the Japan-Korea Basic Scientific Cooperation Program between Korea Science and Engineering Foundation (KOSEF) and Japan Society for the Promotion of Science (JSPS). This research is based on data collected at Bohyunsan Optical Astronomy Observatory (BOAO) that is operated by Korea Astronomy and Space Science Institute (KASI) and Okayama Astrophysical Observatory (OAO) that is operated by National Astronomical Observatory of Japan (NAOJ). We gratefully acknowledge the support from the staff members of BOAO and OAO during the observations. BCL acknowledges the support from the Astrophysical Research Center for the Structure and Evolution of the Cosmos (ARCSEC, Sejong University) of the Korea Science and Engineering Foundation (KOSEF) through the Science Research Center (SRC) program. Data analysis was in part carried out on gsbh computer system operated by the Astronomical Data Center (ADC) and Subaru Telescope of NAOJ. This research has made use of the SIMBAD database, operated at CDS, Stransbourg, France.


\bigskip

\clearpage

\clearpage

\begin{table}
  \begin{center}
    \caption{Stellar parameters of the host star HD 119445}
    \label{Tab:SP}
    \begin{tabular}{cc}
      \hline
      \multicolumn{1}{c}{Parameter} & Value \\
      \hline
       Spectral Type & G6III \\
       $V$ & 6.30 \\
       $B-V$ & 0.879 $\pm$ 0.004 \\
       $\pi$ (mas) & 3.46 $\pm$ 0.71\\
       $M_{v}$ & $-$1.03 \\
       $B.C.$ & $-$0.23 \\
       $T_{\mathrm{eff}}$ (K) & 5083 $\pm$ 103 \\
       $L$ ($\LO$) & 251 $\pm$ 95\\
       $M$ ($\MO$) & 3.9 $\pm$ 0.4\\
       $R$ ($\RO$) & 20.5 $\pm$ 9.3\\
       log $g$ & 2.40 $\pm$ 0.17\\
       $V_{t}$ (km s$^{-1}$) & 1.49 $\pm$ 0.20\\
       $[$Fe/H$]$ & 0.04 $\pm$ 0.18\\
       $v \mathrm{sin} i_{s}$ (km s$^{-1}$) & 6.0 $\pm$ 0.6\footnotemark[$*$], 6.9 $\pm$ 1.0\footnotemark[$\dagger$] \\
      \hline
\multicolumn{2}{@{}l@{}}{\hbox to 0pt{\parbox{85mm}{\footnotesize
\footnotemark[$*$]\citet{Gray1989}\\
\footnotemark[$\dagger$]\citet{deMedeiros1999}
}\hss}}
    \end{tabular}
  \end{center}
\end{table}

\begin{table}
  \begin{center}
    \caption{Radial velocities of HD 119445 at BOAO}
    \label{tab:RVvBOAO}
    \begin{tabular}{ccc}
      \hline
      \multicolumn{1}{c}{JD} & Radial velocity & Error\\
      ($-$2450000) & (m s$^{-1}$) & (m s$^{-1}$) \\
      \hline
3427.3555 	&	215.5 	&	11.6 	\\
3730.3782 	&	$-$309.5 	&	10.6 	\\
3756.3179 	&	$-$265.1 	&	14.2 	\\
3809.2236 	&	96.3 	&	9.9 	\\
3831.2419 	&	199.7 	&	11.7 	\\
3888.0653 	&	457.1 	&	13.3 	\\
4081.3558 	&	$-$253.9 	&	12.6 	\\
4123.2218 	&	$-$361.9 	&	14.3 	\\
4224.1335 	&	94.5 	&	15.2 	\\
      \hline
    \end{tabular}
  \end{center}
\end{table}

\begin{table}
  \begin{center}
    \caption{Radial velocities of HD 119445 at OAO}
    \label{tab:RVvOAO}
    \begin{tabular}{ccc}
      \hline
      \multicolumn{1}{c}{JD} & Radial velocity & Error\\
      ($-$2450000) & (m s$^{-1}$) & (m s$^{-1}$) \\
      \hline
3412.2214 	&	119.9 	&	10.5 	\\
3429.1852 	&	208.0 	&	12.6 	\\
3461.2184 	&	334.6 	&	6.3 	\\
3519.0496 	&	451.6 	&	11.8 	\\
3581.0147 	&	295.5 	&	8.9 	\\
3598.9982 	&	184.2 	&	8.0 	\\
3607.9722 	&	123.7 	&	6.8 	\\
3611.9573 	&	88.0 	&	10.3 	\\
3614.9515 	&	87.7 	&	7.1 	\\
3644.9092 	&	$-$123.8 	&	10.6 	\\
3695.3348 	&	$-$366.1 	&	8.0 	\\
3709.3600 	&	$-$373.1 	&	8.9 	\\
3730.3545 	&	$-$377.2 	&	11.7 	\\
3743.3350 	&	$-$330.3 	&	6.8 	\\
3805.1203 	&	0.4 	&	10.7 	\\
3824.1283 	&	116.4 	&	8.4 	\\
3831.1425 	&	138.5 	&	9.7 	\\
3931.1190 	&	401.2 	&	12.9 	\\
3947.0120 	&	423.0 	&	10.3 	\\
4075.2883 	&	$-$232.4 	&	8.7 	\\
4131.2517 	&	$-$385.6 	&	7.4 	\\
4151.2616 	&	$-$343.3 	&	7.0 	\\
4176.2334 	&	$-$225.9 	&	9.2 	\\
4199.2797 	&	$-$96.4 	&	8.1 	\\
4214.1939 	&	$-$4.2 	&	12.8 	\\
4243.0711 	&	166.6 	&	6.6 	\\
4261.9704 	&	262.2 	&	8.0 	\\
      \hline
    \end{tabular}
  \end{center}
\end{table}

\begin{table}
  \begin{center}
    \caption{Orbital parameters of HD 119445 b}
    \label{tab:OPstar}
    \begin{tabular}{cc}
      \hline
      \multicolumn{1}{c}{Parameter} & Value\\
      \hline
       $K_{1}$ (m s$^{-1}$) & 413.5 $\pm$ 2.6 \\
       $P$ (days) & 410.2 $\pm$ 0.6 \\
       $e$ & 0.082 $\pm$ 0.007 \\
       $\omega$ (deg) & 160.5 $\pm$ 4.3 \\
       $T$ (JD) & 2452873.9 $\pm$ 5.6 \\
       $\Delta$RV\footnotemark[$*$] (m s$^{-1}$) & $-$43.0 $\pm$ 5.1 \\
       rms (m s$^{-1}$) & 13.7 \\
       Reduced $\sqrt{\chi^{2}}$ & 1.7 \\
       $N_{obs}$ & 36 \\
       $a_{1} \mathrm{sin} i_{p}$ (10$^{-3}$AU) & 15.54 $\pm$ 0.10 \\
       $f_{1}$($m$) (10$^{-7}M_{\odot}$) & 29.75 $\pm$ 0.58 \\
       $M_{\mathrm{2}} \mathrm{sin} i_{p}$ ($M_{\mathrm{J}}$) & 37.6 $\pm$ 2.6 \\
       $a$ (AU) & 1.71 $\pm$ 0.06 \\
       \hline
\multicolumn{1}{@{}l@{}}{\hbox to 0pt{\parbox{85mm}{\footnotesize
\footnotemark[$*$]Offset between BOAO and OAO velocities.
}\hss}}
      \end{tabular}
  \end{center}
\end{table}

\end{document}